# In Vivo Quantification of Glioma-Induced Solid Stress Using MR Elastography and Deformable Image Registration

Noah Jaitner[1], Mehrgan Shahryari[1,2], Jakob Schattenfroh[1], Tom Meyer[1], Hossein S. Aghamiry[1], Jakob Ludwig[1], Jakob Jordan[1], Anastasia Janas[2,3,4], Anna Morr[1], Biru Huang[1], Boshra Shams[3,5], Thomas Picht[3,5], Gueliz Acker[3,4], Tobias Schaeffter[6,7], Jing Guo[1], Ingolf Sack[1]

[1]Department of Radiology, Charité – Universitätsmedizin Berlin, Germany

[2]Berlin Institute of Health at Charité – Universitätsmedizin Berlin, Germany

[3]Department of Neurosurgery, Charité – Universitätsmedizin Berlin, Germany

[4]Department of Radiation Oncology and Radiotherapy, Charité – Universitätsmedizin Berlin, Germany

[5]Cluster of Excellence: "Matters of Activity. Image Space Material", Humboldt University, Berlin, Germany

[6]Physikalisch-Technische Bundesanstalt, Berlin, Berlin, Germany

[7]Technical University Berlin, Einstein Center Digital Future, Berlin, Germany

# Abstract

Solid stress is increasingly being recognized as a key driver of tumor progression and aggressiveness, yet it has not been directly measured in patients so far. Here, we combine multifrequency magnetic resonance elastography with 3D magnetic resonance imaging (MRI)-based diffeomorphic deformable image registration network analysis to noninvasively quantify glioma-induced solid stress. In both a mouse model and patients, we identified spatially heterogeneous deformation patterns extending well beyond tumor margins. While deformation magnitude was not found to correlate with tumor size or clinical outcome, excess solid stress - defined as the product of peritumoral volumetric strain and stiffness differential between unaffected brain and peritumoral tissue - was inversely associated with patient survival, highlighting its potential as a quantitative, imaging-derived biomarker. To our knowledge, this study provides the first direct quantification of mechanical stress in patients with glioma.

# Introduction

Gliomas are the most common malignant brain tumors with glioblastoma (GBM) being the most prevalent and lethal subtype, accounting for 54% of newly diagnosed gliomas[1]. GBM has a three-year survival rate of only 3-5% [2] and an incidence of 4.2 per 100,000 person-years [3]. Its poor prognosis is attributed to highly invasive growth, molecular heterogeneity, therapy resistance, and the lack of reliable imaging markers for differentiating true tumor progression from pseudoprogression[4, 5].

Understanding the biophysical and mechanical properties of gliomas, particularly in vivo, is essential, as these influence tumor-host interactions[6], invasion[7, 8], and treatment response[9, 10]. In vivo magnetic resonance elastography (MRE)[11] has revealed that gliomas exhibit unusual soft-solid properties compared to the typically highly viscous brain parenchyma[12-15]. This biomechanical signature has been suggested as a promoter of unstable physical boundary conditions and viscous fingering[16]. In contrast to most extracranial tumors, which share stiffer and more viscous properties than their surroundings, glioma present a reversed mechanical gradient which likely contributes to unique deformation patterns observed in peritumoral regions[17]. Radiomic analyses have consistently shown that peritumoral tissue in glioma exhibits large, heterogeneous deformation amplitudes[18, 19] and peritumoral compression[20], indicating the accumulation of solid stress generated by the growing tumor. Solid stress – defined as the sum of mechanical shear and compressive forces acting on tissue[21] - has been suggested as a key driver of tumor aggressiveness by influencing cell signaling[22], promoting cell unjamming[23, 24], impairing perfusion through vascular compression and worsen neurological outcomes[25]. Therefore, quantifying solid stress in vivo may provide critical insights into glioma progression and therapeutic resistance[26].

However, directly measuring tumor solid stress in patients is difficult as spatially resolved force fields are typically inaccessible through non-invasive imaging techniques[27-29]. Nevertheless, non-linear mechanical deformation models can predict solid stress by multiplying deformation with changes in elastic modulus[30]. Pioneering studies have assessed tumor-related deformation in patients with GBM using image registration to normative brain atlases[19]. However, these approaches lack critical information on the intrinsic mechanical properties of tissue, particularly the shear modulus, since deformation alone is insufficient to quantify solid stress[28]. Conversely, isolated MRE measurement of glioma shear modulus offers limited predictive value for individual patient outcomes[12, 31].

Therefore, we developed a method to quantify mechanical stress in peritumoral tissue induced by both volumetric and shear strain, through the combination of deformation and shear modulus measurements. We introduce a metric for tissue degradation, termed *excess solid stress,* derived from volumetric strain and the differential shear modulus between peritumoral regions and unaffected brain tissue. Our approach combines advanced multifrequency MRE with three-dimensional (3D) strain field analysis using diffeomorphic deformable image registration in patients with glioma. Following validation of 3D strain mapping in a longitudinal mouse model of GBM, we prospectively enrolled glioma patients for combined 3D MRI and MRE scanning. Excess solid stress was then evaluated to investigate its potential as a prognostic imaging biomarker of glioma aggressiveness, with respect to patient survival.

# Methods

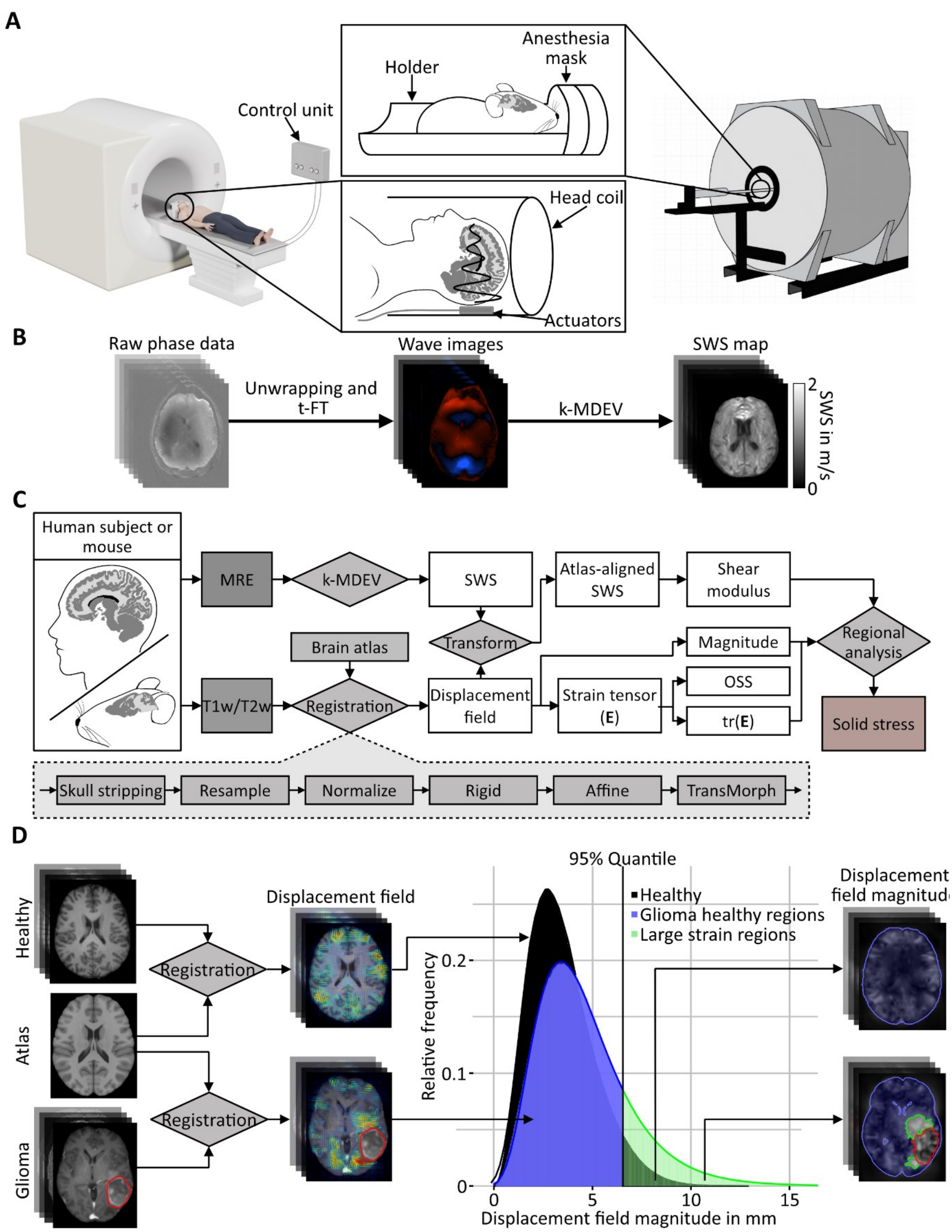

**Figure 1** *Pipeline for the acquisition and analysis of magnetic resonance elastography (MRE) and 3D magnetic resonance imaging (MRI) data in humans and mice.* ***(A)*** *Experimental setup. In human subjects, MRE and 3D-MRI were performed using a 3T clinical MRI scanner at a single time point. Multifrequency MRE vibrations were excited using two compressed-air-powered actuators placed beneath the head. In mice, T2-weighted (T2w) MRI was performed at two time points during tumor progression. Mice were secured in a holder and scanned in a 7T preclinical MRI system.* ***(B)*** *Postprocessing steps of MRE. Raw phase data is first unwrapped, followed by a temporal Fourier transform (t-FT) to extract the fundamental of angular drive frequency $\omega$. The resulting wave images are then inverted according to*

*equations (1) and (2) using k multifrequency dual elasto-visco inversion (k-MDEV).* ***(C)*** *Analysis workflow. Human T1-weighted (T1w) and mouse T2w images, along with corresponding MRE data for humans, were processed. MRE data were inverted using the k-MDEV algorithm to generate shear wave speed (SWS) maps. Anatomical images were registered to species-specific atlases (MNI ICBM Human Brain Atlas[32]; DSURQE Mouse Brain Atlas[33]) by skull stripping, resampling, normalization, rigid and affine transformations, and deformable registration using a transformer-based model. Resulting deformation fields were used to align SWS maps of humans, and the shear modulus was computed. Strain tensors derived from displacement fields enabled calculation of displacement magnitude, volumetric strain ($tr(\boldsymbol{E})$), and octahedral shear strain ($OSS$) according to equations (5), (7a) and (7b). Regions with elevated displacement magnitudes were used to estimate stress (equations (10)-(12)).* ***(D)*** *Regional analysis. Displacement magnitudes were quantified following registration. Histogram analysis was used to compare values across healthy subjects and glioma patients. Regions exceeding the 95th percentile of displacement magnitudes in unaffected tissue were identified and are visualized in green.*

## Study Design

Seven female C57BL/6 mice (10 ± 2 weeks old) were investigated at 13 and 22 days following GBM implantation as previously published in Janas et al.[34]. An additional group of twelve female C57BL/6 mice (12 weeks old) were used as healthy controls. All animal experiments followed German animal protection law and were monitored by local authorities (LaGeSo, Berlin, Germany, registration numbers G0130/20 and G0284/13[34, 35]). Clinical MRI and MRE was performed in twenty-one patients with glioma prior to surgical resection (6 women, 52 ± 12 years, 2 WHO Grade II, 2 WHO Grade III, 17 WHO Grade IV) and 16 healthy volunteers (4 women, 45 ± 13 years, age range: 28 to 60 years). The study was approved by the internal review board of Charité - Universitätsmedizin Berlin (EA1/248/19). Written informed consent was obtained from all participants.

## Specimen preparation

Tumor cell implantation was performed as previously described[34, 36]. In brief, anesthetized mice were placed in a fixation frame, and the skull was exposed. An injection of 20,000 GL261 tumor cells was administered into the right striatum (2 mm lateral, 1 mm anterior, and 3 mm deep from the bregma) using a 1 µL Hamilton syringe over a 15-minute period. Mice were anesthetized with ketamine/xylazine (ketamine-hydrochloride + 1 mg xylazine per 100 g bodyweight) during the procedure.

## Imaging

### *Volumetric MRI in mice*

Imaging was performed on a preclinical 7T MRI scanner (BioSpec 70/20 USR, Bruker, Germany and Paravision 6.0.1 Software) using a 20-mm diameter volume coil (RAPID Biomedical, Rimpar, Germany). The mice were anesthetized with isoflurane (1.0–1.5%, CP Pharma, Burgdorf, Germany) diluted in a mixture of 30% O2 and 70% N2O and placed inside the scanner (see Figure 1A). To assess tumor volume in GBM-bearing mice, 32 contiguous axial image slices were acquired using a T2-weighted (T2w) rapid acquisition with refocused echoes (RARE) sequence (repetition time (TR) = 4200 ms, echo time (TE) = 36 ms, field of view (FoV) = 19.2 x 19.2 mm$^2$, matrix size = 192 x 192, resolution = 0.1 x 0.1 x 0.5 mm$^3$). Nine contiguous slices were acquired in healthy mice (TR = 1800 ms, TE = 54 ms, FoV = 20 x 20 mm$^2$, matrix size = 256 x 256, resolution = 0.078 x 0.078 x 1 mm$^3$).

### *Volumetric MRI and MRE in humans*

Imaging was performed on a clinical 3T MRI scanner (Magnetom Lumina, Siemens, Erlangen, Germany) using a 32-channel head coil. Axial T1-weighted (T1w) images were acquired in a central slap through the brain using the scanner’s auto align function and a magnetization-prepared rapid acquisition gradient echo (MPRAGE) sequence (TR = 2300 ms, TE = 2.27 ms, FoV = 192 × 192 mm$^2$ and matrix size = 128 × 128, resolution = 1.5 x 1.5 x 2 mm$^3$). In glioma patients, either 15 or 36 slices were acquired to minimize acquisition time, depending on the patient's compliance. In healthy subjects, the full brain was covered with 256 slices and a resolution of 1 x 1 x 1 $mm^3$.

Multifrequency MRE was performed using a single-shot, spin-echo, echo-planar imaging sequence with four consecutively applied drive frequencies of 20, 25, 30, and 40 Hz. The central MRE slice was aligned with the central slice of the T1w image block. As shown in Figure 1A, each subject's head was positioned on 3D-printed air cushions[37]. Three orthogonal wave displacement components were encoded using a single-cycle, flow-compensated motion-encoding gradient (amplitude = 34 mT/m, duration = 28 ms). Corresponding to the T1w scans, either 15 or 36 contiguous transverse slices were acquired depending on patient comfort level while 9 slices were acquired in healthy subjects using the same sequence (TR = 4700 ms, TE = 70 ms, FoV = 202 × 202 $mm^2$ and matrix size = 128 × 128, resolution = 1.603 x 1.603 x 2 $mm^3$)

## Data postprocessing

The data postprocessing workflow is presented in Figure 1C. Multifrequency MRE inversion was performed using Matlab 24.2.0.2833386 (R2024b), Update 4 (MathWorks, Natick, MA). Statistical analysis was done in R version 4.4.2. All other postprocessing steps were implemented in Python (Python 3.10.13, Python Software Foundation, USA).

### *Shear modulus calculation*

Shear wave speed (SWS) maps for both mice and human subjects were generated using the k wave-number-based multifrequency dual elasto-visco (k-MDEV) inversion method, which is publicly accessible at https://bioqic-apps.charite.de[38] and outlined in Figure 1B. Briefly, the raw phase data was unwrapped, followed by a temporal Fourier transform to extract the fundamental of angular drive frequency $\omega$ in each voxel. The resulting complex-valued wave images were then directionally filtered along the main axes and diagonal axes, resulting in a set of 8×3×4 shear wave images per slice, $u_{dcf}$, corresponding to eight directions (*d*), three wavefield components (*c*) and three drive frequencies (*f*)[38, 39]. Spatial phase gradients were then used to derive wave numbers

$$k_{dcf} = \|\nabla \arg(u_{dcf})\|. \quad (1)$$

where $\|\cdot\|$ denotes the Euclidean norm, $\nabla$ is the gradient operator, and arg extracts the phase of a complex number. The resulting 8×3×4 $k_{dcf}$ images were compounded into a single $SWS$ map after weighting reciprocal wave numbers with weights $W = |u_{dcf}|^2$ and averaging them across components, directions, and *N* frequencies:

$$\mathrm{SWS} = \frac{N}{\sum_{f=1}^{N} \frac{1}{\mathrm{SWS}_f}} \text{ where } \mathrm{SWS}_f = \frac{\omega \sum_d \sum_c W}{\sum_d \sum_c k_{dcf} W} \quad (2)$$

Finally, $SWS$ maps were converted into shear modulus $\mu$ maps by assuming a unit tissue mass density $\rho$ of 1000 kg/m³ [40] according to

$$\mu = \mathrm{SWS}^2 \rho. \quad (3)$$

Tissue values were distinguished from fluid regions by thresholding SWS maps with 1 m/s[40, 41].

### *Registration*

First image pixel of the skull were removed on human T1w images using SynthStrip[42] - a deep-learning-based tool to strip non-brain tissues from MRI scans for subsequent analysis. Skull pixels in T2w brain images of mice were removed using RS2-NET[43]. The images were then resampled to an isotropic voxel resolution of 1 $mm^3$ and normalized into an intensity range between 0 and 1. Next, rigid and affine registrations were performed, using ANTs[44] to align anatomical structures. Finally, diffeomorphic deformable image registration was applied using TransMorph[45]. A publicly available, pretrained model was used for human subjects[45]. For mouse data, a separate network with the same underlying architecture was trained on 350 mouse datasets. Human images were registered to the MNI ICBM Human Brain Atlas[32] while mouse brains were registered to the DSURQE Mouse Brain Atlas[33], providing the displacement field. Additionally, mouse brains imaged at time point 22 days were registered to their corresponding images acquired at the time point of 13 days.

### *Strain*

The following notations were used for volumetric and shear strain. Computed displacement field $\mathbf{u}(\mathbf{r})$ maps undeformed coordinates $\mathbf{r}$ onto deformed coordinates $\mathbf{r}'$:

$$\mathbf{r}' = \mathbf{r} + \mathbf{u}(\mathbf{r}). \quad (4)$$

The magnitude of displacement is defined by the Euclidean norm

$$D(\mathbf{u}(\mathbf{r})) = \|\mathbf{u}(\mathbf{r})\| = \sqrt{\mathrm{u_1}^2 + \mathrm{u_2}^2 + \mathrm{u_3}^2} \quad (5)$$

The finite strain (large deformation) induced by the growing tumor was defined by the Green-Lagrange strain tensor based on the deformation gradient $\mathbf{F} = \frac{\partial \mathbf{r}'}{\partial \mathbf{r}} = \mathbf{I} + \nabla\mathbf{u}$:

$$\mathbf{E} = \frac{1}{2}(\mathbf{F}^T\mathbf{F} - \mathbf{I}) = \frac{1}{2}(\nabla\mathbf{u} + \nabla\mathbf{u}^T + (\nabla\mathbf{u})^T\nabla\mathbf{u}), \quad (6)$$

where $\mathbf{I}$ is the identity matrix and $\mathbf{u}^T$ denotes the transpose of $\mathbf{u}$. Volumetric strain was defined by the first invariant of the strain tensor $\mathbf{E}$ and the octahedral shear strain (OSS)[46], respectively as:

Volumetric strain: $tr(\mathbf{E}) = E_{11} + E_{22} + E_{33}$ (7a)

Octahedral Shear strain (OSS):

$$\frac{2}{3}\sqrt{(E_{11} - E_{22})^2 + (E_{11} - E_{33})^2 + (E_{22} - E_{33})^2 + 6(E_{12}^2 + E_{13}^2 + E_{23}^2)} \quad (7b)$$

Note, the sign of volumetric strain indicates the direction of deformation with negative strain referring to compression and positive strain expansion. In contrast, OSS has only positive values. Since image registration to the pre-deformed, non-tumoral reference state is not feasible in patients, we used a mouse model of GBM tumors for validation and variability analysis. This enabled us to assess differences between atlas-based and individual reference-based registrations and to thus identify potential biases in strain estimation.

### *Regional Analysis*

To account for interindividual anatomical variability, regional deformation analysis was performed. Displacement field magnitude according to equation (5) was computed within a cohort of healthy controls. The 95th percentile was used as the threshold below which deformation was considered nonsignificant in patients. Conversely, regions above this threshold were classified as abnormally deformed regions and were observed around tumors (Figure 1D). To mitigate high displacement magnitudes due to boundary artifacts, we restricted analysis to a brain mask eroded by five voxels and disregarded regions smaller than 10 voxels. Deviations of individual brain anatomy from normative brain atlas anatomy were accounted for by subtracting group-mean healthy volumetric strain $\overline{tr(\boldsymbol{E})_0}$ and mean healthy $\overline{OSS_0}$ from each patient's strain:

$$\overline{tr(\mathbf{E})} = tr(\mathbf{E})_{glioma} - \overline{tr(\mathbf{E})_0}, \quad (8a)$$

$$\overline{OSS} = OSS_{glioma} - \overline{OSS_0}. \quad (8b)$$

### *Excess solid stress*

For isotropic elastic materials, Hooke's law relates strain to stress. However, when dealing with finite (medium-sized) deformations, we must consider geometric nonlinear effects, and the infinitesimal strain tensor used in Hooke's law is replaced by the nonlinear Green–Lagrange strain tensor $\mathbf{E}$, as defined in equation (6). The resulting constitutive law defines the second Piola-Kirchhoff stress tensor

$$\mathbf{\Sigma} = \lambda\, tr(\mathbf{E})\mathbf{I} + 2\mu\mathbf{E} \tag{9}$$

with $\lambda$ and $\mu$ denoting the first and second Lamé's parameter, respectively. $\mu$ is also referred to as the shear modulus. Volumetric stress, also known as pressure, is given by

$$\mathbf{\Sigma}_{vol} = -\frac{\mathbf{1}}{\mathbf{3}} tr(\mathbf{\Sigma}) = -(\frac{2\nu}{1-2\nu} + \frac{2}{3})\mu \cdot tr(\mathbf{E}). \tag{10}$$

In this equation, $\lambda$ was substituted with $\frac{2\nu\mu}{1-2\nu}$ , where $\nu$ is Poisson's ratio – a measure of compressibility, which is 0.5 for incompressible materials like water. Because brain tissue has a high water content of 70–80%, it is generally considered nearly incompressible, resulting in values of $\nu$ close to 0.5[47]. However, solid tumors deform surrounding tissue over longer time scales, inducing fluid drainage, tissue atrophy, and necrosis. As a result, mass conservation is violated, leading to volumetric changes in peritumoral areas. Consequently, tumor-deformed tissue may exhibit substantially increased compressibility, with reported Poisson's ratios on the order of 0.4[48-51]. Unlike volumetric stress, which is pressure, shear stress is associated with volume-conserving deformations, represented by the off-diagonal elements of the Green–Lagrange strain tensor $\mathbf{E}$. These elements can be combined into the shear-strain invariant, OSS[46], as defined in Equation (7b), which quantifies the maximum shear strain in a deformed solid. Analogously, we define maximum shear stress as

$$\Sigma_{shear} = 2\mu \cdot OSS. \tag{11}$$

Tumor-induced solid stress, $\Sigma = \Sigma_{vol} + \Sigma_{shear}$, acts on surrounding tissue over extended time periods, leading to tissue damage and remodeling. While volumetric changes are reflected in the Lamé parameter $\lambda$ and Poisson's ratio $\nu$, structural tissue changes are reflected in shear modulus $\mu$. Since structural changes, such as tissue degradation associated with softening, are potentially crucial for prognosis and outcome, we introduce the differential shear modulus $\Delta\mu = \mu_0 - \mu_1$, where $\mu_0$ and $\mu_1$ denote the shear moduli before and after deformation, respectively. In clinical practice, the reference state of $\mu_0$ before deformation is unknown. As an approximation, we assume that apparently unaffected brain tissue in patients with glioma represents the pre-deformation state with shear modulus $\mu_0$, while $\mu_1$ corresponds to the peritumoral region deformed by the tumor. Then, the product of differential shear modulus $\Delta\mu = \mu_0 - \mu_1$ (reflecting the mechanical consequence) and strain (reflecting the mechanical cause) provides a quantitative measure of *excess solid stress* $\sigma_{vol}$ and $\sigma_{shear}$ corresponding to the type of deformation, i.e., volumetric (vol) or shear:

$$\sigma_{\mathrm{vol}} = \Delta\mu \cdot tr(\mathbf{E}) \tag{12a}$$

$$\sigma_{\mathrm{shear}} = \Delta\mu \cdot OSS \tag{12b}$$

Statistical Analysis

Nonparametric tests were used for all statistical analyses to avoid bias induced by the small sample size[52, 53]. A Wilcoxon signed-rank test was used to test for differences in volumetric strain, $OSS$, $SWS$, and stress between different regions in the brain. A Wilcoxon-Mann-Whitney test was applied to test for differences in volumetric strain, $OSS$, and $SWS$ between glioma patients and healthy subjects. The relationship between strain and stiffness as well as excess solid stress and survival was determined using linear regression. No censoring was performed. Pearson's linear correlation coefficient R along with p-values compared to the constant model are reported. Group values are reported as mean and standard deviation (SD). p-values <0.05 were considered significant.

# Results

*Different registration methods consistently measure glioblastoma-induced displacement in mice*

Figure 2A presents the displacement field from a reference brain state in two representative mice, derived from image registration to either a standard anatomical atlas or earlier time points. In both cases, pronounced outward displacement is observed from the tumor into the surrounding brain tissue. The direction of this displacement remained consistent whether the mouse brains were registered to a standard anatomical atlas or to reference time points. Tumors exhibited a mean volumetric increase of 863 ± 100 % between time point 13 days and 22 days as previously analyzed[34]. Tumor size at time point 22 days relative to the mouse brain atlas was $34 \pm 20$ % (range: 12 % to 68 %). Figure 2B shows representative T2w images, displacement field magnitude, volumetric strain, and $OSS$ maps of the brain of two mice for atlas and reference time point registration. For healthy control mice, mean displacement field magnitude was $1.83 \pm 0.04\ mm$ with a 95th quantile of $3.22\ \mathrm{mm}$. In contrast, GBM mice registered to the atlas exhibited a significantly higher mean displacement field magnitude of $2.6 \pm 0.5\ mm$ compared to healthy controls (p < 0.001). For GBM mice registered to a reference time point at 13 days, the average displacement field magnitude was $3.4 \pm 2.0\ mm$. Regions exceeding the 95th percentile, defined as abnormal deviations from a standard brain showed perimeter locations and size consistently across registrations to atlas and reference time point.

*Glioblastoma in mice induces volumetric strain*

Group statistical analysis of normalized volumetric strain, $\overline{tr(\mathbf{E})}$, and normalized shear strain, $\overline{OSS}$, for healthy, atlas-registered GBM, and reference time point-registered GBM mice are shown in Figure 2C. In healthy control mice, $\overline{tr(\mathbf{E})}$ was $-0.0 \pm 0.5$ % while $\overline{OSS}$ was $0.0 \pm 0.7$ %. For atlas-registered GBM mice, $\overline{tr(\mathbf{E})}$ was significantly lower than in healthy controls, both for unaffected brain regions ($-9 \pm 1$ %, p < 0.001) and in peritumoral regions ($-9 \pm 2$ %, p < 0.001). $\overline{OSS}$ for atlas-registered GBM mice was $0.7 \pm 1.1$ % for unaffected brain regions and $-0.5 \pm 2.5$ % in peritumoral regions. Similarly, in reference time point-registered GBM mice, $\overline{tr(\mathbf{E})}$ was significantly lower than in healthy controls for both unaffected brain regions ($-5 \pm 2$ %, p < 0.001) and peritumoral regions ($-5 \pm 5$ %, p = 0.0098). In contrast, $\overline{OSS}$ was similar in unaffected brain regions ($-0.9 \pm 3.6$ %) and peritumoral regions ($-0.3 \pm 4.0$ %), indicating that GBM in the mouse brain primarily induces volumetric strain rather than shear strain. Volumetric strain was lower when computed using atlas-based registration compared to the registration to a reference time point. The mean difference between the two registration methods was 3%, which we identify as a systematic error associated with volumetric strain estimation. Conversely, $\overline{OSS}$ did not differ significantly between atlas-based and reference time point-based registration.

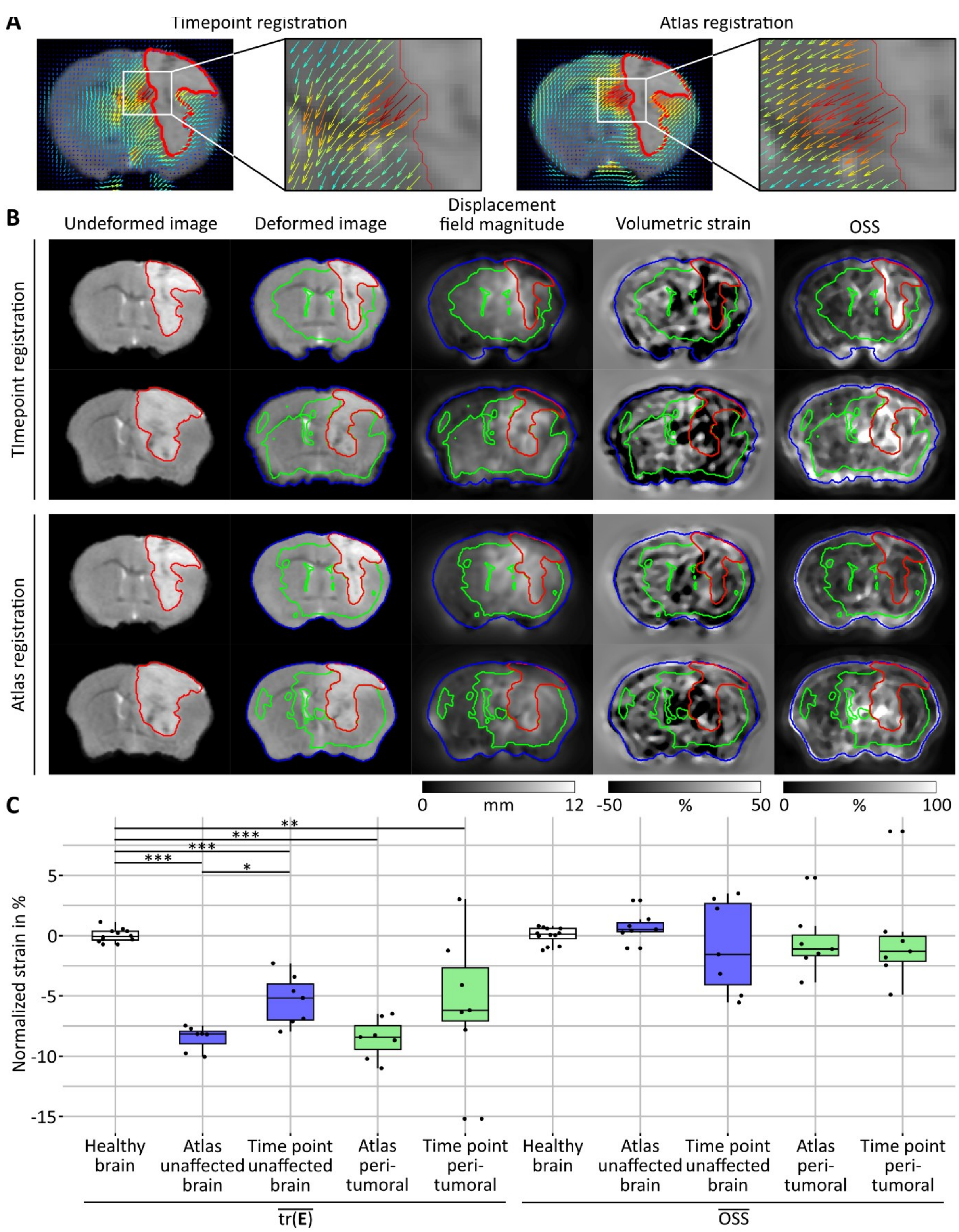

**Figure 2** *Results for in vivo mouse brain.* **(A)** *Representative displacement fields for two distinct mouse brains with magnified insets. The tumor region is outlined in red.* **(B)** *Results for two GBM mice with image registration to reference time point and to standardized atlas. Shown are the undeformed image, deformed image, displacement magnitude, volumetric strain, and OSS. Peritumoral regions with high displacement magnitude are outlined in green and unaffected regions in blue.* **(C)** *Group statistical analysis of normalized volumetric strain $\overline{tr(\boldsymbol{E})}$ and normalized shear strain $\overline{OSS}$ for healthy mice, atlas-registered GBM mice, and reference time point registered GBM mice (green: large-strain peritumoral region, blue: unaffected brain region). *p<0.05. **p<0.01, ***p<0.001*

<u>*Glioma in patients induces spatially heterogeneous peritumoral displacement*</u>

Figure 3A shows representative displacement fields obtained in two glioma patients. As highlighted in the insets, these fields reveal heterogeneously distributed areas of tissue displacement directed away from the tumor and exerted on the tumor-surrounding regions. Tumor size relative to the human brain atlas was $1.5 \pm 1.5$ % (range: 0.2 % to 5.8 %), which was significantly smaller than in the mouse model (p < 0.001). Figure 3B presents T1w images, $SWS$ maps, displacement field magnitude, volumetric strain, and OSS for the brains of two patients and two healthy volunteers. The magnitude of displacement in healthy volunteers was $3.0 \pm 0.3\ mm$, with a 95[th] percentile value of $6.5\ \mathrm{mm}$. Henceforth the 95[th] percentile of the displacement field magnitude was taken as threshold for abnormal deviations from standard brain anatomy in patients. Regions of abnormal displacement were heterogeneously distributed in the peritumoral regions of patients. No such regions were observed in healthy volunteers. Displacement magnitudes in non-tumor brain were significantly higher in glioma patients than in healthy volunteers ($3.7 \pm 0.5\ mm$, p < 0.001). Regions of abnormal displacement did not overlap with edema (Dice score: 0.10 $\pm 0.089$ ) as further exemplified by representative cases in Supplementary Figure 1.

*Peritumoral regions are highly compressed and are softer than healthy tissue*

As seen in Figure 3B, compressive and shear strain dominate in tumor-surrounding regions, while in healthy subjects, large strain appeared predominantly at brain boundaries. Group statistical results for $\overline{tr(\mathbf{E})}$, $\overline{\mathrm{OSS}}$, and $SWS$ and the corresponding regions are shown in Figure 4A and B, respectively. $\overline{tr(\mathbf{E})}$ was highest, with a negative sign, in peritumoral regions, indicating compression ($-17 \pm 9$ %), followed by unaffected brain regions in patients ($-9 \pm 5$ %, p < 0.001) and healthy brains in volunteers ($0 \pm 2$ %, p < 0.001). $\overline{\mathrm{OSS}}$ was also highest in peritumoral regions ($23 \pm 9$ %), followed by unaffected brain regions in patients ($13 \pm 6$ %, p < 0.001) and healthy volunteers ($0 \pm 2$ %, p < 0.001). Mean SWS in healthy brain tissue was $1.37 \pm 0.06\ m/s$, while SWS in unaffected brain regions ($1.25 \pm 0.03\ m/s$, p < 0.001) and peritumoral tissue of glioma patients ($1.27 \pm 0.03\ m/s$, p < 0.001) was significantly lower. Tumor SWS ($1.21 \pm 0.07\ m/s$) was significantly lower than that of whole brain in healthy controls (p < 0.001), unaffected glioma brain (p = 0.038) and peritumoral regions (p = 0.007).

*Degraded (softer) peritumoral tissue is highly deformed*

Figure 4C shows that shear modulus in peritumoral tissue ($\mu_1$) increased with volumetric strain in peritumoral regions, because $\overline{tr(\mathbf{E})} < 0$ in all patients, consistently indicating compression. (slope: $0.5 \pm 0.2$ kPa, intercept: $1.71 \pm 0.03$ kPa, R = 0.55, p = 0.009). Further, peritumoral shear modulus $\mu_1$ linearly decreased with shear strain $\overline{OSS}$, which, by definition, is always positive (slope: $-0.5 \pm 0.2$ kPa, intercept: $1.75 \pm 0.04$ kPa, R = -0.60, p = 0.004). Thus, larger strain, both volumetric and shear, was associated with softer tissue properties as a sign of deformation-induced degradation of peritumoral tissue. Combined assessment of strain and shear modulus enabled us to determine in vivo solid stress, $\Sigma = \Sigma_{vol} + \Sigma_{shear}$, according to equations (10) and (11). Assuming a Poisson's ratio of $\nu$ = 0.4 according to Moran et al.[49] (see Methods), volumetric solid stress $\Sigma_{vol}$ was $1.2 \pm 0.6\ kPa$ with a range of $2.2 \pm 0.2\ kPa$ to $0.1 \pm 0.2\ Pa$. Significantly lower values were found for shear stress $\Sigma_{shear}$ with $0.7 \pm 0.3\ kPa$ and ranges of $1.1 \pm 0.2\ kPa$ to $0.3 \pm 0.1\ kPa$ (p < 0.001).

*Highly deformed peritumoral tissue is stiffer than apparently unaffected tissue*

Although we observed tissue degradation and softer values of $\mu_1$ in conjunction with higher strain magnitudes in peritumoral regions, differential modulus $\Delta\mu = \mu_0 - \mu_1$ showed the opposite trend of lower values at higher strain, both volumetric (slope: $-0.6 \pm 0.2\ m/s$, intercept: $-0.16 \pm 0.04\ m/s$, R = -0.56, p = 0.008) and shear (slope: $0.7 \pm 0.2\ m/s$, intercept: $-0.21 \pm 0.05\ m/s$, R =

0.63, p = 0.002, Figure 5A). This observation strongly indicates that the whole brain is mechanically altered when glioma-induced tissue compression occurs.

*Excess solid stress in GBM is associated with worse prognosis*

Since $\Delta\mu$ reflects differential stiffness - and thus local (peritumoral) and global tissue (unaffected brain) degradation - we further investigated the prognostic value of excess solid stress, defined as the product of $\Delta\mu$ and strain according to equations (12) and potentially representing the mechanical driver of degradation. To this end, we analyzed a subgroup of 10 patients (1 WHO grade II, 1 WHO grade III, and 8 WHO grade IV) for whom survival data were available (Mean: 836 ± 466 days, range: 162 to 1514 days). Excess solid stress $\sigma_{vol}$ showed a significant linear decline with length of survival (slope: $-0.017 \pm 0.006\ Pa/days$, intercept: $-19\ \pm\ 6\ Pa$, R = -0.70, p = 0.02, Figure 5B). Mean $\sigma_{vol}$ was $4.3 \pm 11.6\ Pa$, ranging from $30 \pm 7\ Pa$ to $-13 \pm 6\ Pa$. Excess solid stress did not correlate with tumor size or WHO grade, due to the limited sample size of grade II and III tumors. Mean $\sigma_{shear}$ was $-7.2 \pm 14.2\ Pa$, ranging from $16 \pm 65\ Pa$ to $-31 \pm 70\ Pa$. In line with our solid stress analysis, which demonstrated that $\Sigma_{vol}$ was approximately twice as high as $\Sigma_{shear}$, $\sigma_{vol}$ showed a stronger association with survival compared to $\sigma_{shear}$ (p = 0.13). Preoperative Karnofsky performance status scale and Aachen Aphasia Test[54] results revealed no significant impairment in 8 of the 10 patients, suggesting no direct correlation between mechanical stress and survival in this subset.

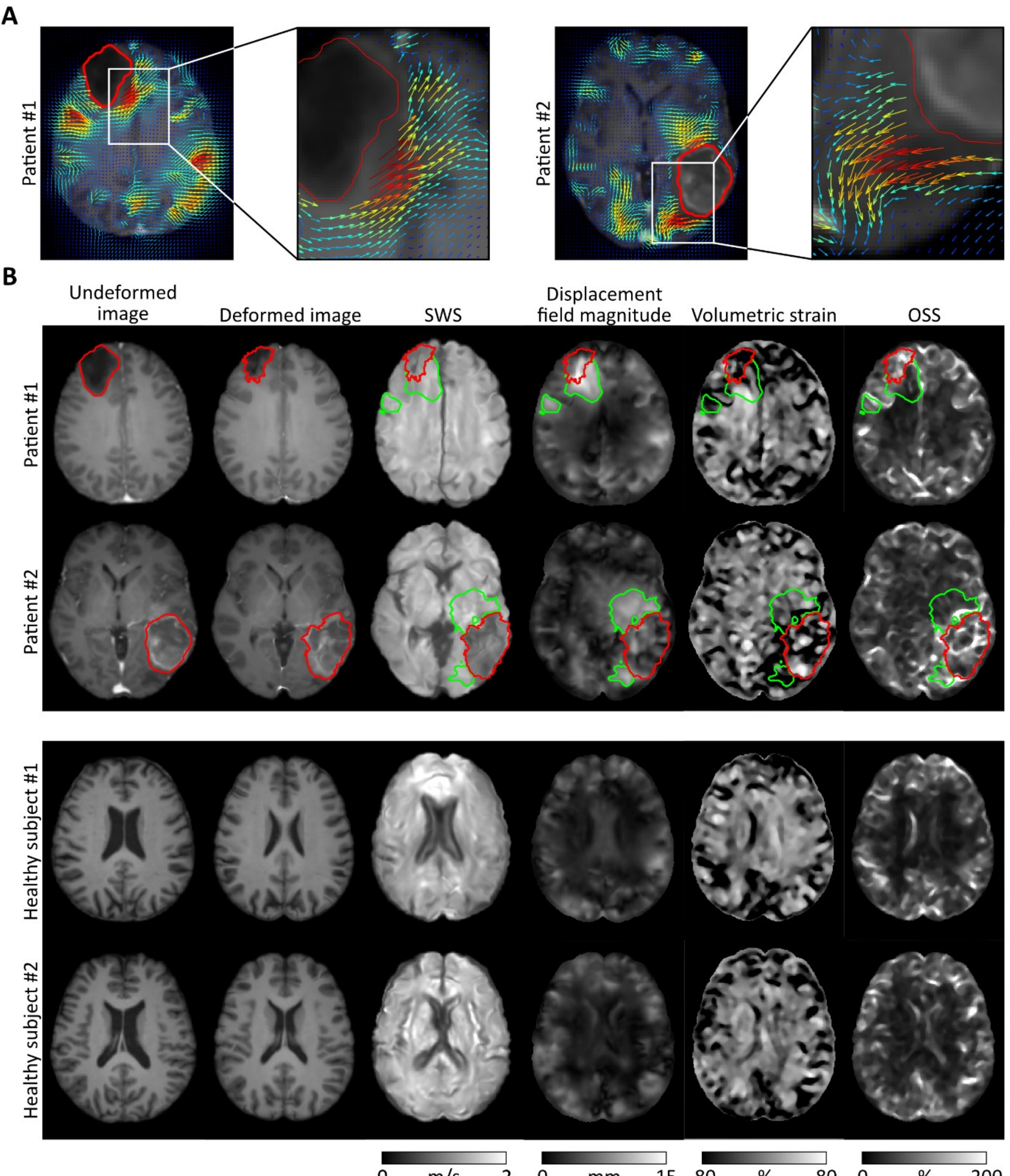

**Figure 3** *Representative result maps illustrating findings in glioma patients and healthy controls.* **(A)** *Displacement fields in a central axial slice from two patients with glioma (Patient #1 – male, 30years, Grade II, IDH-mutant; Patient #2 -female, 58years Grade IV, IDH-wildtype). Tumor boundaries are outlined in red; magnified views highlight regional deformation.* **(B)** *Image registration outputs for two glioma patients and two healthy controls, including undeformed/deformed images, displacement magnitude, volumetric strain, and* OSS*. Peritumoral regions exhibiting abnormally increased displacement magnitudes are outlined in green; tumor margins are shown in red.*

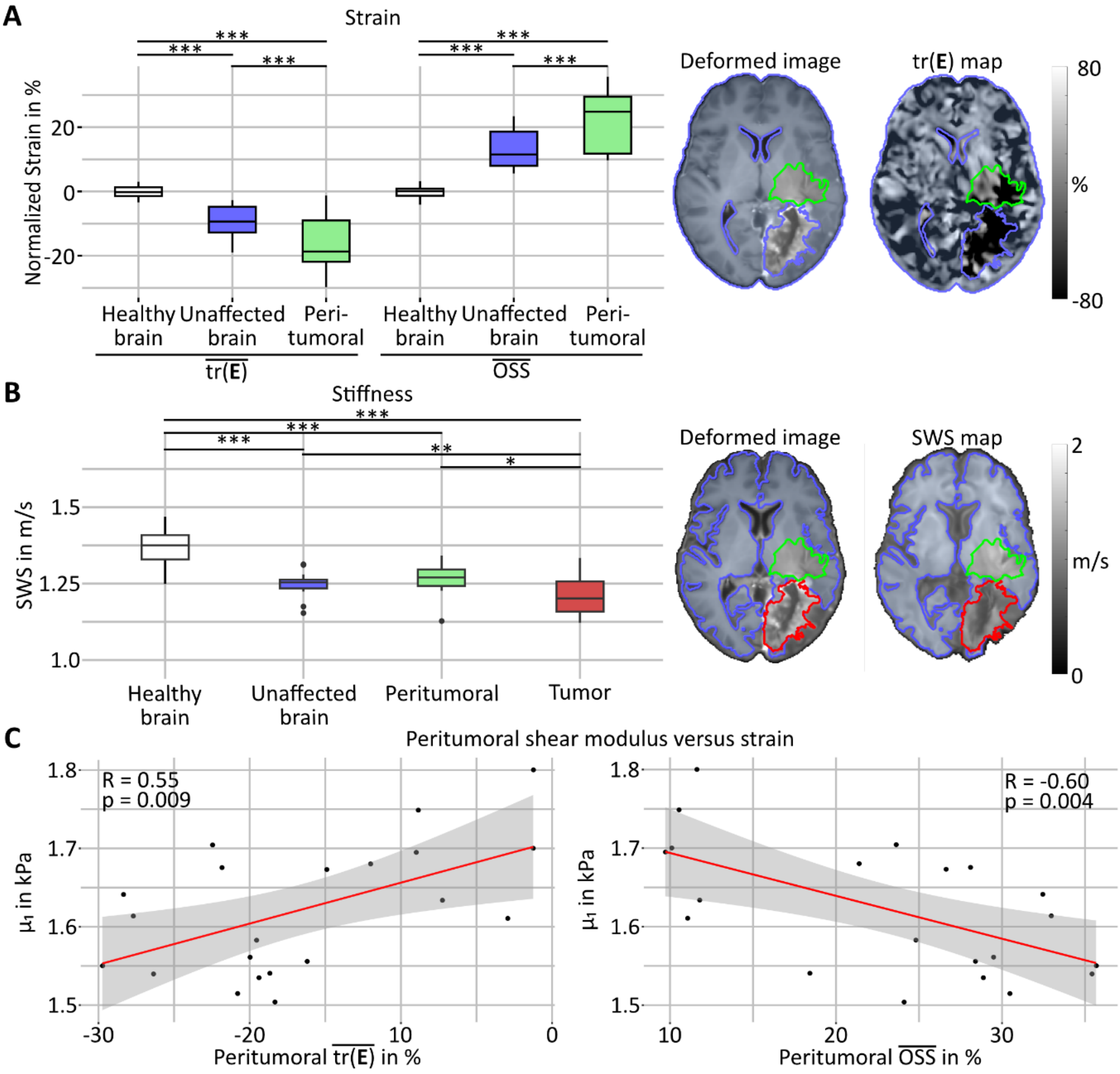

***Figure 4*** *Group statistical results for human subjects.* ***(A)*** *Group comparisons of normalized volumetric strain* $\overline{tr(\boldsymbol{E})}$ *and* $\overline{OSS}$ *between glioma patients and healthy controls across defined regions (green: peritumoral area, blue: unaffected brain region without tumor).* ***(B)*** *Group comparisons of* SWS *across glioma patients and controls, averaged over green (peritumoral), red (tumor), and blue (unaffected) regions as indicated.* ***(C)*** *Scatter plot of peritumoral shear modulus* $\mu_1$ *versus normalized volumetric strain* $\overline{tr(\boldsymbol{E})}$ *and* $\overline{OSS}$ *across glioma patients. Linear regression revealed increasing peritumoral shear modulus with normalized volumetric strain* $\overline{tr(\boldsymbol{E})}$ *(slope:* $0.5 \pm 0.2\ kPa$*, intercept:* $1.71 \pm 0.03\ kPa$*, R = 0.55, p = 0.009) and decreasing with* $\overline{OSS}$ *(slope:* $-0.5 \pm 0.2\ kPa$ *, intercept:* $1.75 \pm 0.04\ kPa$ *, R = -0.60, p = 0.004).* **p<0.05. **p<0.01, ***p<0.001*

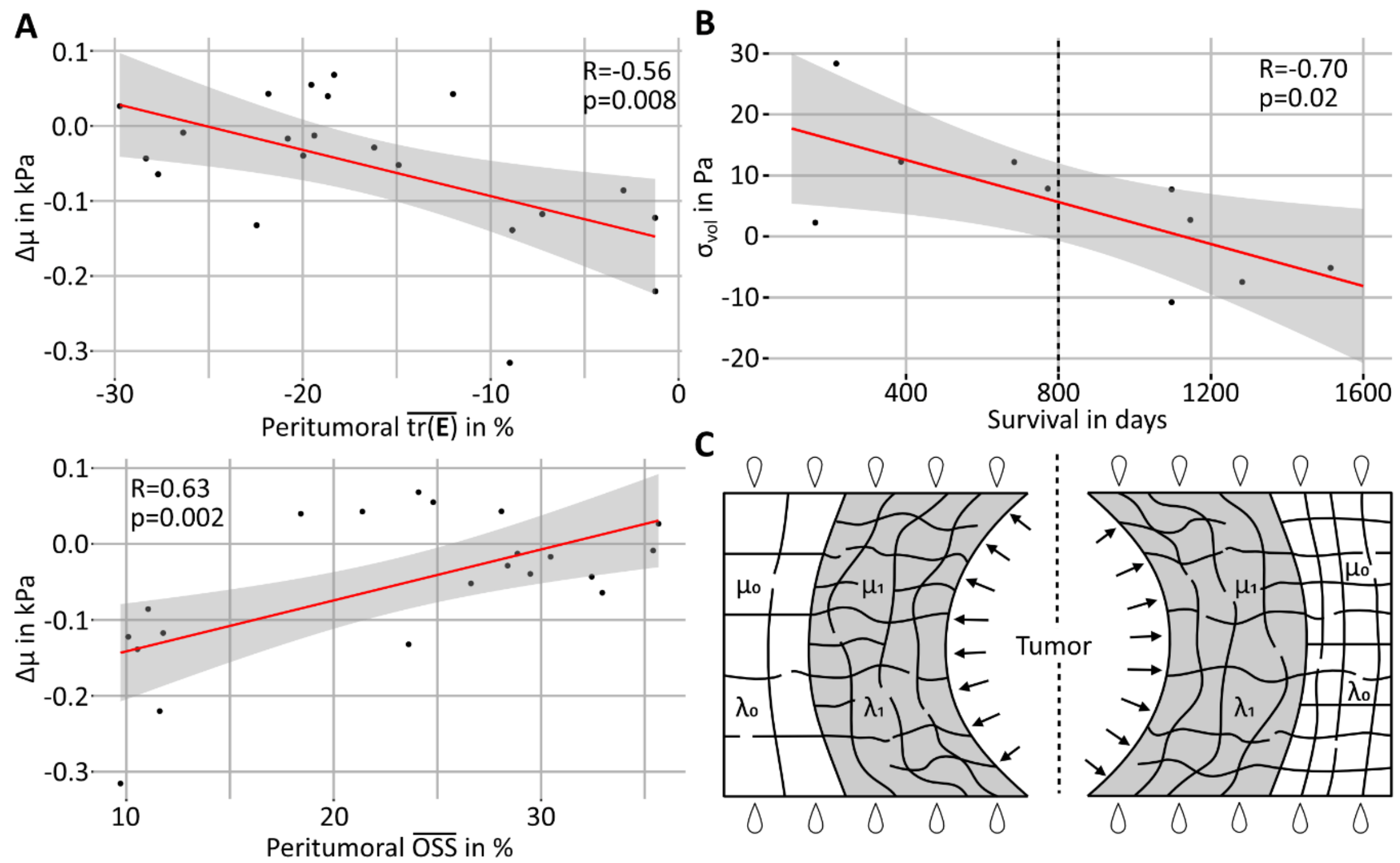

***Figure 5*** *Group statistical results for patients with glioma and mechanical alterations in peritumoral brain tissue.* ***(A)*** *Scatter plot of differential shear modulus $\Delta\mu$ versus normalized volumetric strain $\overline{tr(\boldsymbol{E})}$ and $\overline{OSS}$ across patients. Linear regression revealed decreasing differential shear modulus $\Delta\mu = \mu_0 - \mu_1$ with normalized volumetric strain $\overline{tr(\boldsymbol{E})}$ (slope: $-0.6 \pm 0.2\ m/s$, intercept: $-0.16 \pm 0.04\ m/s$, R = -0.56, p = 0.008) and increasing with $\overline{OSS}$ (slope: $0.7 \pm 0.2\ m/s$, intercept: $-0.21 \pm 0.05\ m/s$, R = 0.63, p = 0.002).* ***(B)*** *Excess volumetric stress $\sigma_{vol}$ plotted against patient survival. A negative correlation was observed (slope = $-0.017 \pm 0.006\ Pa/days$, intercept = $19 \pm 6\ \ Pa$, R = -0.7, p = 0.02), suggesting that lower excess stress improves survival.* ***(C)*** *Schematic illustration of mechanical processes in brain tissue during tumor progression. The peritumoral region is shown in gray. As the tumor expands, it exerts stress on the surrounding tissue, promoting interstitial fluid drainage and tissue degradation. Throughout this process, the first Lamé parameter ($\lambda_0 = \lambda_1$) remains effectively constant, consistent with the tissue's near-incompressibility[30]. In contrast, shear modulus $\mu$ increases in response to compression stiffening[22] or decreases in response to tissue degradation[55]. On the left, the peritumoral region is stiffer compared to unaffected tissue ($\mu_1 > \mu_0, \sigma_{vol} > 0$), while, on the right, this trend is reversed ($\mu_0 > \mu_1$, $\sigma_{vol} < 0$). Stiffer healthy (unaffected) tissue relative to the peritumoral region is associated with a better patient prognosis, as indicated by the dotted line.*

# Discussion

Our study conceptually merges two previously distinct MRI-based approaches, 3D deformation analysis and stiffness mapping, to measure mechanical stress in glioma patients. The novel biomarker, excess stress, integrates cause and consequence of brain tissue degradation from a biomechanical perspective, that is, large deformation and intrinsic softening. Probably for this reason, elevated excess stress correlates with shorter patient survival, suggesting a potential mechanobiological link to glioma aggressiveness and therapeutic responsiveness.

The peritumoral region in glioma is a complex and dynamic microenvironment, characterized by the interplay of cellular and extracellular components that contribute to tumor growth, invasion, and resistance to therapy[6, 56]. GBM exhibits highly infiltrative and irregular growth patterns with striking capacities to establish a brain-wide neuronal circuit connectome[7]. This highly invasive behavior results in substantial displacement from normal brain atlas of brain tissue with a gradient from high displacement to subtle displacements invisible on a coarse MRI grid[57-59]. Areas exhibiting high displacement are heterogeneously distributed around the tumor core, extending deep into apparently healthy brain tissue. These observations are consistent with prior studies demonstrating that glioma infiltration is facilitated in mechanically heterogeneous environments, where stiffness gradients contribute to tumor progression[6, 60]. As a proxy of large deformation, radiographic midline shifts has been reported and correlated with patient survival[61]. Our approach extends this concept by quantifying three-dimensional deformation patterns across the entire brain, revealing a more comprehensive picture of tumor-induced mechanical imbalance. The peritumoral region consistently exhibited higher volumetric strain compared to apparently unaffected brain matter in both mice and humans, indicating the dominating compression force exerted by the tumor. In contrast, shear deformation measured by OSS was significantly elevated only in patients but not in mice. This difference may suggest distinct tumor growth patterns across species and brain sizes, highlighting the critical role of the biological environment and tumor model in solid stress accumulation. For example, the mesenchymal and vascular characteristics of our GL261 glioma model were associated with significantly lower volumetric strain than that observed in patients[25]. This is particularly notable because, relative to total brain volume, tumors were approximately 56% larger in mice than in humans. This difference suggests that tissue replacement rather than displacement occurs during tumor progression in the mouse model.

The observed positive correlation between shear modulus and volumetric strain in glioma patients indicates that mechanically softer peritumoral regions undergo greater compression (larger negative volumetric strain) than their stiffer counterparts. This finding may be explained by chronic compression, which can lead to tissue degradation and softening, similar to the pressure-related stiffness reduction reported in hydrocephalus patients[62]. Together, these observations support a model of mechanical heterogeneity in glioblastoma, characterized by a softened tumor core and surrounding regions of stress accumulation driven by differential growth, cellular proliferation, and confinement. Notably, this compression-induced brain softening arises from large strains acting on the tissue over weeks, months, or years of tumor development and causes tissue-degrading mechanical stress. In contrast, *compression stiffening* of brain tissue, as reported by Janmey and colleagues[22], describes a short-term hyperelastic stiffness response of *intact* tissue.

Rather than hyperelastic stiffening, we observed softening in both peritumoral and unaffected brain tissue of glioma patients compared to healthy volunteers. This finding is consistent with previous reports of glioma-associated tissue remodeling and structural degradation far beyond the tumor margins, reflecting a diffuse, whole-brain process[63-65]. In our study, patients with stiffer peritumoral regions compared to unaffected tissue ($\mu_0 < \mu_1$) had positive excess volumetric stress ($\sigma_{vol} > 0$) and shorter survival than patients where this trend was reversed ($\mu_0 > \mu_1$, $\sigma_{vol} < 0$ ). Therefore, differential stiffness $\Delta\mu$ between large-strain (peritumoral) and low-strain (apparently unaffected) tissue seems to be a parameter of GBM biomechanics. Neither the size of large-strain regions or the tumor, nor any stiffness parameter of tumor or peritumoral region (all being softer than healthy brain tissue) correlated with length of survival. A possible explanation of this association of solid-stress related biomechanics to glioma survival is the interplay between tissue degradation and residual compression stiffening. Patients with widespread degradation (low $\mu_0$) in apparently unaffected brain matter and highly deformed peritumoral tissue (higher $\mu_1$) fall into the poor-prognosis group irrespective of which of the two regions contributes most to this unfavorable scenario. Vice versa, patients with intact global brain matter mechanistically benefit from higher $\mu_0$ values and more favorable biomechanical scenarios. Our method discriminates between pre-stressed intact and degraded tissue and quantifies tissue integrity versus degradation using the combined strain–stiffness parameter of *excess volumetric stress*, highlighting its potential as an imaging-derived biomarker while underscoring the need for validation in larger, controlled studies.

Comparing our results with the literature, we note that our healthy brain stiffness values fall in the range of previously reported values[66, 67]. Furthermore, our findings agree with earlier work demonstrating that tumor tissue is softer than surrounding healthy tissue[12, 31]. Peritumoral regions have not consistently been found to demonstrate stiffening compared to healthy surrounding tissue. Nia et al.[27] reported ex vivo radial solid stress values of approximately $20$ Pa in GBM using a planar-cut method following tissue embedding in agarose, while in situ estimates based on a needle-biopsy technique coupled with a mathematical model yielded circumferential stress values around $100$ Pa. Intraoperative estimates fell into the range of $10$ to $600$ Pa[68]. The shear stress values obtained in our study fall within a similar range. However, the volumetric solid stress values we report are higher, possibly due to differences in methodology and physiological context. A critical parameter in solid stress estimation is the Poisson's ratio, which is not directly measurable in vivo and thus relies on modeling assumptions. Nia et al. employed a relatively low value ($\nu = 0.1$), whereas we adopted a higher value ($\nu = 0.4$), which is closer to the near-incompressibility of brain tissue used in standard models[49]. In incompressible materials, volumetric deformation would require infinite stress, which is unphysical. This highlights the limitations of modeling brain tissue as an elastic solid over longer time periods, where a growing tumor acts as fluid-like mass source [69]. In an unjacked poroelastic scenario, fluid outflow from the tissue surrounding a growing tumor mass mitigates pressure buildup and relaxes volumetric stress. Moreover, GBM not only displaces brain tissue but also gradually replaces it, altering its local mechanical properties and degrading tissue structure.

Although encouraging, our study has limitations. First, the sample size is relatively small and homogeneous, particularly for the analysis of clinical prognosis. Therefore, validation in larger, more heterogeneous populations will be essential to substantiate our findings. Second, our approach relies on registering images to a standardized brain atlas, which can introduce inaccuracies, especially in individuals whose anatomy differs substantially from the norm. To mitigate this, we included healthy control participants and adjusted our displacement measurements to account for the influence of atypical brain anatomy. Furthermore, the registration framework does not account for signal changes due to structural changes. This is particularly relevant in regions affected by edema, where signal changes may not correspond to true anatomical displacement.  Third, despite the use of diffeomorphic registration techniques, the reliance on image contrast may result in misalignments as underlying mechanical properties are not taken into consideration. Future work could address this by incorporating physics-informed registration frameworks that implement stiffness measurements directly into strain estimation. Lastly, in our estimation of volumetric solid stress, we adopted a Poisson's ratio based on values reported in other studies[49]. Performing a joint estimation of the Lamé parameters alongside volumetric stress might allow a more precise quantification of solid stress.

In summary, we present a novel, noninvasive method to quantify volumetric and shear stress in patients with glioma and a mouse GBM model, leveraging deformation mapping and MRE towards prognostic assessment of excess volumetric stress. We demonstrate that high volumetric compression - particularly in softer, compliant brain regions - extends into healthy tissue, indicating heterogeneous stress distributions that mirror glioma infiltration patterns. Our findings demonstrate the significance of mechanical stress as a critical, quantifiable MRI biomarker of glioma pathophysiology and underscore its potential for diagnosis and therapy monitoring.

# Acknowledgements

Support from the German Research Foundation (DFG): CRC1340 Matrix in Vision, CRC1540 EBM, GRK2260 BIOQIC and FOR5628, is gratefully acknowledged. The authors thank the Scientific

Computing Unit of the IT Division of *Charité – Universitätsmedizin Berlin* for providing computational resources that have contributed to the research results reported in this paper. Dr. Shahryari is a participant in the BIH Charité Junior Digital Clinician Scientist Program funded by the Charité – Universitätsmedizin Berlin, and the Berlin Institute of Health at Charité (BIH).

# Data Availability Statement

The data that supports the findings of this study, except clinical imaging data, are available from the corresponding author upon request. Summary statistics and region average values for all subjects are provided in the supplementary material.

# Conflict of Interest

The authors declare no conflict of interest.